\documentclass{aa}
\usepackage{graphicx}

\begin{document}

%\thesaurus{01(02.01.4;                 % atomic processes
%              02.12.1;                 % line: formation
%              03.20.8;                 % techiniques: spectroscopic
%              11.17.1;                 % quasars: absorption lines
%              11.17.4 IRAS 13349+2438; % quasars: individual
%              13.25.2)                 % X-rays: galaxies
%}

\title{Complex resonance absorption structure in the X-ray spectrum of
       \object{IRAS~13349+2438}\thanks{Based on observations obtained with
       {\it XMM-Newton}, an ESA science mission with instruments and
       contributions directly funded by ESA Member States and the USA
       (NASA).}}

\author{M.\,Sako\inst{1}
        \and S.\,M.\,Kahn\inst{1}
        \and E.\,Behar\inst{1}
        \and J.\,S.\,Kaastra\inst{2}
        \and A.\,C.\,Brinkman\inst{2}
        \and Th.\,Boller\inst{3}
        \and E.\,M.\,Puchnarewicz\inst{4}
        \and R.\,Starling\inst{4}
        \and D.\,A.\,Liedahl\inst{5}
        \and J.\,Clavel\inst{6}
        \and M.\,Santos-Lleo\inst{6}}

\institute{Department of Physics and Columbia Astrophysics Laboratory,
           550 West 120th Street, New York, NY 10027, USA
           \and Space Research Organization of the Netherlands,
           Sorbonnelaan 2, 3548 CA, Utrecht, The Netherlands
           \and Max-Planck-Institut fuer Extraterrestrische Physik,
           Postfach 1603, 85741 Garching, Germany
           \and Mullard Space Science Laboratory,
           University College, London, Holmbury St. Mary, Dorking,
           Surrey, RH5 6NT, UK
           \and Physics Department,
           Lawrence Livermore National Laboratory,
           P.O. Box 808, L-41, Livermore,  CA 94550
           \and XMM Science Operations, Astrophysics Division,
           ESA Space Science Dept., P.O. Box 50727,
           28080 Madrid, Spain}

\authorrunning{M. Sako et al.}
\titlerunning{{\it XMM-Newton} Observation of \object{IRAS~13349+2438}}

%\offprints{M. Sako, \email{masao@astro.columbia.edu}}

\date{Received 29 September 2000 / Accepted}

\abstract{The luminous infrared-loud quasar \object{IRAS~13349+2438} was
  observed with the {\it XMM-Newton} Observatory as part of the Performance
  Verification program.  The spectrum obtained by the Reflection Grating
  Spectrometer (RGS) exhibits broad ($v \sim 1400 ~\rm{km~s}^{-1}$ $FW\!H\!M$)
  absorption lines from highly ionized elements including hydrogen- and
  helium-like carbon, nitrogen, oxygen, and neon, and several iron L-shell
  ions (\ion{Fe}{xvii -- xx}).  Also shown in the spectrum is the first
  astrophysical detection of a broad absorption feature around $\lambda = 16 -
  17$ \AA\ identified as an unresolved transition array (UTA) of $2p - 3d$
  inner-shell absorption by iron M-shell ions in a much cooler medium; a
  feature that might be misidentified as an \ion{O}{vii} edge when observed
  with moderate resolution spectrometers.  No absorption edges are clearly
  detected in the spectrum.  We demonstrate that the RGS spectrum of
  \object{IRAS~13349+2438} exhibits absorption lines from at least two
  distinct regions, one of which is tentatively associated with the medium
  that produces the optical/UV reddening.
  }

\maketitle

\section{Introduction}

  \object{IRAS~13349+2438} is an archetypal highly-polarized radio-quiet
  quasar at a redshift of $z = 0.10764$ (Kim et al.\ \cite{kim95}).  Since its
  identification as an infrared-luminous quasar (Beichman et al.\
  \cite{beichman86}), this source has been extensively studied in the optical,
  infrared, and X-ray bands.  In a detailed investigation of the optical and
  infrared spectra and polarization measurements, Wills et al.\
  (\cite{wills92}) demonstrated that the nuclear spectrum exhibits two
  distinct components; a highly-reddened component and a highly-polarized
  component that suffers much lower extinction.  Based on these observational
  facts, Wills (\cite{wills92}) constructed a simple and elegant picture of
  the nuclear region of \object{IRAS~13349+2438} in which the direct AGN
  radiation is attenutated by a thick dusty torus, while the observed
  highly-polarized light is produced by scattering in an extended bipolar
  region, either by warm electrons or by small dust grains.

  \object{IRAS~13349+2438} was detected in the {\it ROSAT} All-Sky-Survey
  (Walter \& Fink \cite{walter93}; Brinkmann \& Siebert \cite{brinkmann94}),
  and has been the target of extensive pointed PSPC observations with {\it
  ROSAT} (Brandt, Fabian, \& Pounds \cite{brandt96}), and with {\it ASCA}
  (Brinkmann et al.\ \cite{brinkmann96}; Brandt et al.\ \cite{brandt97}).  The
  soft X-ray spectrum obtained with the PSPC shows a lack of absorption by
  cold material indicated by the observed optical reddening, and suggests the
  presence of a warm, dusty medium along the line of sight (Brandt, Fabian, \&
  Pounds \cite{brandt96}).

  In a more recent investigation of the complex X-ray properties of
  \object{IRAS~13349+2438}, Siebert, Komossa, \& Brinkmann (\cite{siebert99})
  self-consistently accounted for the effects of dust embedded in the warm
  absorbing medium, and concluded that single zone models, both with and
  without internal dust, do not provide adequate fits to the combined, {\it
  ROSAT}, {\it ASCA}, and optical data sets.  In particular, they find that a
  dust-free warm absorber model formally gives the best fit to the X-ray data,
  and conclude that the X-ray absorption and optical reddening must arise in
  spatially distinct regions.  However, owing to the moderate
  spectral-resolving-power capabilities of the available detectors on {\it
  ROSAT} and {\it ASCA}, and the likely cross-calibration uncertainties, the
  precise nature of the soft X-ray spectrum has remained controversial.

  In this Letter, we present results from the first high-resolution X-ray
  observation of \object{IRAS~13349+2438} with the {\it XMM-Newton}
  Observatory.  The spectrum obtained with the Reflection Grating Spectrometer
  (RGS) shows a wealth of discrete spectral features, including the first
  astrophysical detection of inner-shell $2p -3d$ absorption by M-shell iron
  ions in the form of an unresolved transition array (UTA).  From a detailed
  analysis of the rich absorption spectrum, we measure the column density and
  velocity field of the line-of-sight material.  We show that the spectrum
  contains absorption features from regions with two distinct levels of
  ionization.  The column density of the lower ionization component is
  consistent with the observed optical reddening, and we tentatively associate
  this component with the dusty torus.

\section{Observation and Data Reduction}

  \object{IRAS~13349+2438} was observed with the {\it XMM-Newton} observatory
  (Jansen et al.\ \cite{jansen01}) on 19 -- 20 June, 2000 during the
  Performance Verification phase for a total exposure time of 42~ks.  The data
  obtained with the Reflection Grating Spectrometer (RGS; den Herder et al.\
  \cite{denherder01}) were filtered through standard event-selection criteria
  using the {\it XMM-Newton} Science Analysis Software (SAS).  The source
  spectrum was extracted using a spatial filter in dispersion/cross-dispersion
  coordinates to isolate \object{IRAS~13349+2438} from other possible
  contaminating sources and to reduce contribution from background events.
  Subsequently, the first order events were selected by applying a
  dispersion/pulse-height filter.  The background spectrum was generated using
  all of the events that lie outside the spatial mask.  Wavelengths were then
  assigned to the dispersion coordinates.  The current wavelength scale is
  accurate to within $\sim 8$~m\AA\ across the entire RGS band of $\lambda = 5
  - 35$~\AA\ ($E = 0.35 - 2.5$~keV).

  The European Photon Imaging Camera (EPIC; Turner et al.\ \cite{turner01};
  St\"{u}der et al.\ \cite{struder01}) MOS1 data were also processed with the
  SAS.  The MOS2 detector was operated in FAST UNCOMP mode, for which
  reduction by the SAS is not possible as yet.  Source events were extracted
  using a circular region of radius 45\arcsec.  A nearby source-free region
  with a radius of 3\arcmin\ was used to assess the background.

  Three modes of the Optical Monitor (OM; Mason et al.\ \cite{mason01}) were
  used during the observation.  The V grism (or optical grism; exposure time
  3000~s), UV grism (exposure time 1000~s), and the UVW2 filter ($\sim 1500 -
  3000$~\AA; effective exposure time 6340~s).  There is no significant
  variability in the UVW2 observations.  Absolute flux calibrations for the
  grism data were not available at the time of writing.  The optical spectrum
  shows clear evidence of H$\beta$ emission but the signal-to-noise is
  insufficient to determine the presence of any other emission lines (within
  an observer frame wavelength range of $\sim 3000 - 6000$~\AA). The UV
  spectrum (coverage $\sim 2000 - 3500$~\AA) is too weak for spectral
  extraction.

\section{Results of Spectral Fitting}

\subsection{Underlying Continuum Radiation}

  To first obtain a rough characterization of the shape of the continuum, we
  use the PN spectrum, which has the highest statistical quality and covers a
  broad range in energy.  During the observation, the source exhibited a
  gradual and steady increase in brightness by $\sim 30$\%, but with no
  noticable change in the spectral shape.  We, therefore, use the cumulative
  spectrum for all of our subsequent analyses.  The $0.2 - 10$ keV spectral
  region can be well-fitted with a phenomenological model that consists of a
  powerlaw and two black body components absorbed through a Galactic column
  density of $N_{\rm{H}}^{\rm{gal}} = 1.1 \times 10^{20} ~\rm{cm}^{-2}$
  (Murphy et al.\ \cite{murphy96}).  The PN data require two black body
  components with temperatures of $kT \sim 70$ eV and $\sim 12$ eV.  The
  best-fit powerlaw photon index of $\Gamma \sim 2.2$ is consistent with the
  value implied by the {\it ASCA} data (Brinkmann et al.\ \cite{brinkmann96};
  Brandt et al.\ \cite{brandt97}).

  The MOS1 data from $0.3$ to $10$ keV, excluding the $0.6 - 1.2$ keV region
  where the RGS shows complex absorption features, were best-fit using a
  single blackbody plus power-law (a $\chi_r^2$ of 1.59 for 199 degrees of
  freedom), which is generally consistent with the PN data.  The blackbody
  temperature is $kT \sim 110$ eV.  The best-fit photon index is $\Gamma =
  2.1$ with a column density of $1.4 \times 10^{20} ~\rm{cm}^{-2}$, which is
  only slightly higher than the Galactic value.

\begin{figure*}
  \resizebox{18cm}{!}{\rotatebox{-90}{\includegraphics{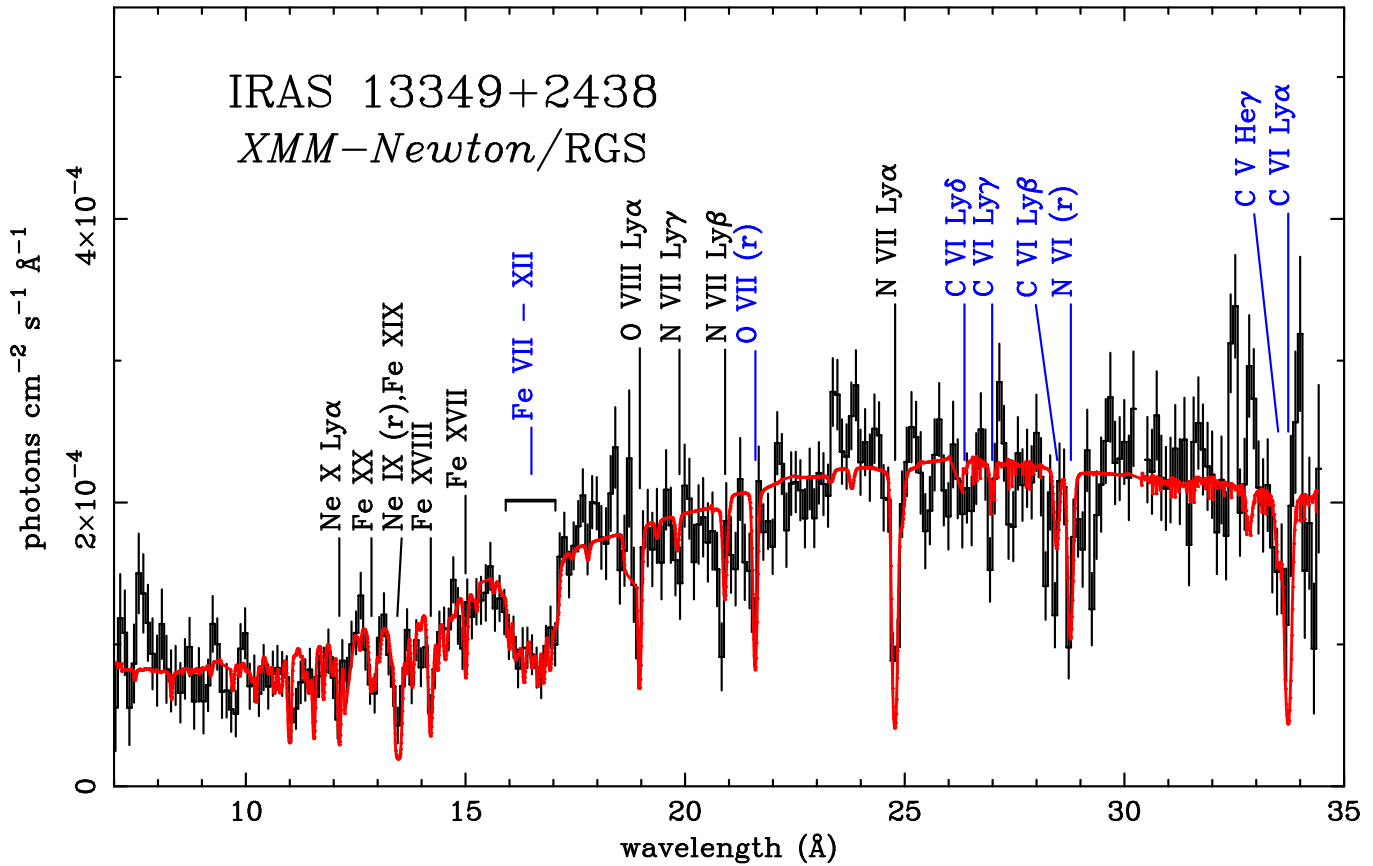}}}
  \caption{The RGS first order spectrum of \object{IRAS~13349+2438} corrected
           for cosmological redshift ($z = 0.10764$).  The error bars
           represent $1 \sigma$ Poisson fluctuations.  The wavelength bins
           are approximately $\sim 0.1$~\AA\ wide.  The best-fit model
           spectrum is superimposed in red.  Absorption line features labeled
           in blue are predominantly produced in the low-$\xi$ component,
           while those labeled in black originate from the high-$\xi$
           component (see text for full details).  The discrepancy at
           $\lambda \sim 29.3$ \AA\ is probably due to inner-shell absorption
           by \ion{N}{v} (lithium-like), which is not included in our model.}
  \label{fig:f1}
\end{figure*}

\subsection{The Absorption Spectrum of \object{IRAS~13349+2438}}

  The RGS spectrum shown in Figure~\ref{fig:f1} exhibits numerous absorption
  lines from a wide range in levels of ionization.  The most prominent
  features in the spectrum are K-shell absorption lines of H- and He-like
  carbon, nitrogen, oxygen, and neon, and L-shell lines of \ion{Fe}{xvii --
  xx}.  The spectrum also shows a broad absorption feature between $\lambda
  \sim 16 - 17$~\AA\ ($E \sim 730 - 770$~eV).  The observed location of the
  red ``edge'' of this feature is at $\lambda = 17.10 \pm 0.05$~\AA\ in the
  rest-frame of the quasar, and is close but undoubtedly inconsistent with
  that of the \ion{O}{vii} photoelectric edge ($\lambda = 16.78$~\AA).  The
  shape of the absorption trough towards the shorter wavelengths is also
  incompatible with that of a photoelectric edge.  We identify this feature as
  a UTA of inner-shell $2p -3d$ resonance absorption lines in relatively
  cool, M-shell iron.  The shape of this feature is strikingly similar to
  laboratory absorption measurements of a heated iron foil (Chenais-Popovics
  et al.\ \cite{chenais00}), and also agrees well with our own calculations as
  described below.

  We adopt a continuum model similar to that inferred from the PN data; i.e.,
  the sum of a powerlaw and a blackbody component, the latter merely in order
  to parametrize empirically the soft X-ray spectral shape.  We also fix the
  powerlaw photon index to $\Gamma = 2.2$ and the column density of cold
  material to the Galactic value.  With the continuum model defined, we then
  apply absorption components of H- and He-like C, N, O, and Ne, and
  \ion{Fe}{xvii -- xxiv}.  Each ion is treated as a separate component in the
  spectral fit and contains all relevant resonance transtions from the
  ground-state and photoelectric edges in the RGS band.  The line profiles are
  calculated accounting for thermal and turbulent velocity broadening.
  Transition wavelengths and oscillator strengths were calculated with the
  atomic physics package HULLAC (Bar-Shalom et al.\ \cite{barshalom98}),
  except for the wavelengths of the strong Fe~L resonance lines where we use
  laboratory measurement values described in Brown et al.\ (\cite{brown00}).
  We use photoionization cross sections from Verner et al.\ (\cite{verner96}).
  For the iron UTA absorption, photoexcitation cross sections from L-shell to
  M-shell in \ion{Fe}{v -- xvi} are computed.  All of the transitions $2l^8$
  $3l^x$ - $2l^7$ $3l^{x+1}$ ($x$ = 1 through 12) are taken into account, of
  which the $2p - 3d$ excitations are the most important.  For the atomic
  structure, the most significant configuration mixings, which conserve the
  total angular momentum within the $n = 3$ shell (namely $3p^2 + 3s3d$), are
  included.  This approximation is expected to be adequate for analysing the
  presently observed unresolved absorption feature.  A more detailed
  discussion of the UTA is presented in Behar, Sako, \& Kahn (\cite{behar01}).

  Each component is then convolved through the instrument line spread function
  and fit for the ion column density simultaneously with the black body
  continuum parameters and the normalization of the powerlaw component.  The
  best-fit black body temperature is $kT \sim 100$ eV with a flux in the $5 -
  35$~\AA\ range of $4.0 \times 10^{-3} ~\rm{photons~cm}^{-2} ~\rm{s}^{-1}$.
  The flux in the powerlaw component is $2.4 \times 10^{-3}
  ~\rm{photons~cm}^{-2} ~\rm{s}^{-1}$ in the same wavelength range.

  The observed widths of the absorption lines, as shown below, are much larger
  than both the instrument line spread function and the thermal broadening of
  a gas with $kT \sim 10 ~\rm{eV}$, which is the expected temperature for a
  photoionized plasma at this level of ionization.  With the current
  statistical quality of the RGS data, however, we are not able to constrain
  the turbulent velocities $v_{\rm{turb}}$ of the individual ion components.
  We, therefore, assume a uniform mean turbulent velocity field, keeping in
  mind that each ion can, in principle, exist in regions of different
  turbulent velocities.  The derived ion column densities, therefore, may be
  uncertain to some degree, as quantified in the following section.

  We obtain a statistically acceptable fit to the RGS data with $\chi^2_r =
  1.07$ for 532 degrees of freedom.  The continuum parameters inferred from
  the RGS data are also consistent with those derived from PN and MOS.  The
  intrinsic isotropic luminosities in the 0.3 -- 2~keV and 2 -- 10~keV regions
  are $L_X \sim 2 \times 10^{44} ~\rm{erg~s}^{-1}$ and $\sim 5 \times 10^{43}
  ~\rm{erg~s}^{-1}$ (assuming $H_0 = 65 ~\rm{km~s}^{-1} ~\rm{Mpc}^{-1}$ and
  $q_0 = 0.5$), respectively, which are lower than both the {\it ROSAT} PSPC
  and {\it ASCA} values by a factor of $\sim 5$.  The measured ion column
  densities are listed in Table~\ref{tbl:t1}.  To illustrate the statistical
  significance of the various components, we also list the changes in $\chi^2$
  when each of the components is removed from the model.  The best-fit
  $FW\!H\!M$ turbulent velocity is $v_{\rm{turb}} = 1430^{+360}_{-280}
  ~\rm{km~s}^{-1}$, which is much larger than the thermal velocity of a
  photoionized medium.  We also find weak evidence of an average bulk outflow
  velocity shift with $v_{\rm{shift}} = +200^{+170}_{-180} ~\rm{km~s}^{-1}$,
  where positive velocity denotes a blueshift.

\begin{table}[htbp]
  \begin{center}
    \caption[Measured $N_i$]{\label{tbl:t1} Measured Ion Column Densities}
    \begin{tabular}{clrc}
      \\ \hline \hline
      Ion & $N_i$ (cm$^{-2}$) $^a$ & $\Delta \chi^2$ $^b$ &
       Component$^c$ \\ \hline \hline

      C V     &  $(6.3^{+5.2}_{-4.8}) \times 10^{16}$ &   4.8 & low \\
      C VI    &  $(6.0^{+6.4}_{-3.1}) \times 10^{16}$ &  23.7 & low \\
      N VI    &  $(2.4^{+1.4}_{-1.0}) \times 10^{16}$ &  29.1 & low \\
      N VII   &  $(1.3^{+0.8}_{-0.4}) \times 10^{17}$ & 109.5 & low/high \\
      O VII   &  $(3.7^{+2.7}_{-1.8}) \times 10^{16}$ &  24.2 & low \\
      O VIII  &  $(9.5^{+8.7}_{-4.7}) \times 10^{16}$ &  24.7 & high \\
      Ne IX   &  $(1.2^{+0.6}_{-0.5}) \times 10^{18}$ &   9.9 & high \\
      Ne X    &  $(4.9^{+10}_{-3.2}) \times 10^{17}$  &  25.9 & high \\
      Fe XVII  &  $(1.7^{+1.6}_{-1.2}) \times 10^{17}$ &  23.0 & high\\
      Fe XVIII &  $(6.3^{+2.4}_{-1.9}) \times 10^{17}$ & 115.0 & high \\
      Fe XIX  &  $(9.7^{+4.8}_{-3.9}) \times 10^{17}$ &   166.7 & high \\
      Fe XX   &  $(3.0^{+2.8}_{-2.0}) \times 10^{17}$ &    39.5 & high \\
              &  & \\
      Fe VII  &  $(1.5^{+1.5}_{-1.3}) \times 10^{16}$ &  9.6 & low \\
      Fe VIII &  $(4.6^{+1.5}_{-1.3}) \times 10^{16}$ & 91.8 & low \\
      Fe IX   &  $(8.8^{+12}_{-8.7}) \times 10^{15}$  & 29.3 & low \\
      Fe X    &  $(2.4^{+1.3}_{-1.1}) \times 10^{16}$ & 72.4 & low \\
      Fe XI   &  $(1.9^{+1.1}_{-0.9}) \times 10^{16}$ & 65.3 & low \\
      Fe XII  &  $(6.4^{+10}_{-4.4}) \times 10^{16}$  & 25.6 & low \\ \hline

    \end{tabular}\\[1.0ex]
  \end{center}
  $^a$ The turbulent velocities for all of the ions are fixed to the same
       value.  We find a best-fit with $v_{\rm{turb}} = 1430^{+360}_{-280}
       ~\rm{km~s}^{-1}$ $FW\!H\!M$. \\
  $^b$ The increase in $\chi^2$ when the ion component is excluded from the
       best-fit model. \\
  $^c$ The dominant $\xi$-component responsible for producing the absorption
       feature (see Figure~\ref{fig:f1} and text for details).
\end{table}

\section{Implications of the Results of Spectral Fitting}

  For the measured ion column densities listed in Table~\ref{tbl:t1}, many of
  the strong absorption lines of neon and iron L ions are in the logarithmic
  region of the curves of growth, which indicates that the derived column
  densities are highly coupled with the assumed turbulent velocity.  Those of
  K-shell carbon, nitrogen, and oxygen, and iron M-shell ions, however, are in
  the linear regime, and the derived column densities are not very sensitive
  to the turbulent velocity.  An important point to note is that the observed
  velocity widths may as well be due to a superposition of multiple, discrete
  absorption components each of which are optically thin, and are unresolved
  with the RGS.  Such features have been observed in UV absorption line
  spectra of many Seyfert 1 galaxies (Crenshaw et al.\ \cite{crenshaw99}, and
  references therein), some of which show as many as 7 distinct, kinematic
  components (e.g., Mrk 509; Kriss et al.\ \cite{kriss00}).  If this is the
  case, the measured column densities from the neon and iron L lines, which
  are produced in the logarithmic region of the curve of growth, may be
  underestimated relative to the true value.  The column densities of the
  He-like species, as well as those of the UTA, are not affected by this,
  since the total observed values are still in the optically thin regime.

  As shown in Table~\ref{tbl:t1}, the detections of absorption lines from
  \ion{Fe}{vii -- xii} and \ion{Fe}{xvii -- xx} are highly significant.  The
  column densities of the intermediate charge states of \ion{Fe}{xiii -- xvi},
  however, are consistent with zero, with an upper limit of $N_i \sim \rm{few}
  \times 10^{15} ~\rm{cm}^{-2}$ for each of these charge states.  This
  indicates that the line-of-sight material consists of either a multi-phase
  gas in a single medium, or two (or more) spatially distinct regions.
  Motivated by this observational fact, we refit the spectrum using the same
  model, except, we assume that the line-of-sight material consists of two
  discrete velocity components; (1) the low-ionization-parameter component
  including \ion{C}{vi}, \ion{C}{v}, \ion{N}{vi}, \ion{O}{vii}, and the
  M-shell Fe, and (2) the high-ionization-parameter component including
  \ion{N}{vii}, \ion{O}{viii}, \ion{Ne}{ix}, \ion{Ne}{x}, and L-shell Fe.
  Contrary to our previous fit where the bulk velocities of all the ions were
  fixed relative to one another, we find that the low-ionization-parameter
  component is significantly blue-shifted with $v_{\rm{shift,low}} =
  +420^{+190}_{-180} ~\rm{km~s}^{-1}$, while the bulk velocity shift in the
  high-ionization component is consistent with zero ($v_{\rm{shift,high}} =
  -20^{+200}_{-330} ~\rm{km~s}^{-1}$).  This implies that the low-ionization
  gas is being accelerated substantially compared to the high-ionization
  component, as would be expected in a radiatively-driven outflow (Arav \& Li
  \cite{arav94}).  The derived turbulent velocities of both components remain
  consistent with that of our previous fit ($v_{\rm{turb}} \sim 1500
  ~\rm{km~s}^{-1}$ $FW\!H\!M$).

  From the observed distribution of charge states of M-shell iron, the average
  ionization parameter, $\xi = L/nr^2$ ($\rm{erg~cm~s}^{-1}$), of the
  absorbing gas is estimated to be $\log \xi \sim 0$ based on a calculation
  with the photoionization code XSTAR (Kallman \& Krolik \cite{kallman95})
  using the inferred continuum shape for the ionizing spectrum.  The width of
  the distribution in ionization parameter is no larger than $\Delta \log \xi
  \sim 1$.  The measured iron ion column densities suggest that the
  corresponding equivalent hydrogen column density is $N_{\rm{H}} \sim (1 - 3)
  \times 10^{21} ~\rm{cm}^{-2}$ assuming a solar iron abundance.  This
  low-$\xi$ gas accounts for most of the carbon and He-like nitrogen and
  oxygen absorption lines as well, however, with some indication of lower
  carbon and nitrogen abundances by a factor of $\sim 2$ and oxygen by a
  factor of $\sim 3$ relative to the solar iron abundance.  The lack of
  absorption from \ion{Fe}{xiii -- xvi}, however, indicates that substantial
  amounts of material with ionization parameters in the range $1 \la \log \xi
  \la 2$ are not present along the line-of-sight.  Whether this is related to
  the global structure of the circumnuclear medium or a mere coincidence from
  a superposition of physically distinct regions is not known.

  The absorption lines from H-like nitrogen and oxygen, H- and He-like neon,
  and L-shell iron are produced in a medium with $2.0 \la \log \xi \la 2.5$
  and an equivalent hydrogen column density of $N_{\rm{H}} \sim (1 - 4) \times
  10^{22} ~\rm{cm}^{-2}$, again, assuming a solar iron abundance.  The
  inferred abundances of nitrogen, oxygen, and neon relative to that of iron
  are consistent with solar values, but are not well-constrained.

  For a normal dust-to-gas ratio, the observed reddening of $E(B-V) = 0.3$ in
  \object{IRAS~13349+2438} (Wills et al.\ \cite{wills92}) corresponds to a
  hydrogen column density of $N_{\rm{H}} = 1.7 \times 10^{21} ~\rm{cm}^{-2}$
  (Burstein \& Heiles \cite{burstein78}).  This value is significantly lower
  than the total amount of X-ray absorbing material (low-$\xi$ + high-$\xi$)
  observed in the present X-ray spectrum.  Coincidentally, however, the
  derived column density of the low-$\xi$ component is very close to that of
  the optical reddening, although, we cannot conclusively associate the
  low-$\xi$ X-ray absorber with the dusty torus.  The column density of the
  high-$\xi$ component, on the other hand, is a factor of $\sim 10$ higher.

  An interesting point to note is that the $1 \la \log \xi \la 2$ region is
  {\it not} thermally unstable, based on XSTAR calculations described above.
  On the other hand, the high-$\xi$ region ($2.0 \la \log \xi \la 2.5$) that
  we observe in the spectrum {\it is} thermally unstable.  However,
  complications such as non-solar metal abundances and/or inaccurate
  ionization and recombination rates may alter the shape of the thermal
  stability curve significantly, and, hence, the temperature ranges of the
  unstable regions (Hess, Kahn, \& Paerels \cite{hess97}; Savin et al.\
  \cite{savin99}).

  For the observed X-ray luminosity of $L_X \sim 2.5 \times 10^{44}
  ~\rm{erg~s}^{-1}$, the high-$\xi$ component with $\log \xi \ga 2.0$ provides
  the constraint, $n_e r^2 \la 2.5 \times 10^{42} ~\rm{cm}^{-1}$.  The
  measured column density through this medium is $N_H \sim n_e \Delta r \ga
  10^{22} ~\rm{cm}^{-2}$, where $\Delta r$ is the radial thickness.  Assuming
  that $\Delta r/r < 1$, these constraints combined provide an upper limit in
  the location of the high-$\xi$ gas of $r_{\rm{high-}\xi} < 2.5 \times
  10^{20} ~\rm{cm} \sim 80 ~\rm{pc}$, which is representative of a typical
  narrow-line region.  The correponding gas density in this region is $n_e \ga
  40 ~\rm{cm}^{-3}$ with an estimated thickness of $\Delta r \sim 80 ~\rm{pc}
  \sim r_{\rm{high-}\xi}$.  A similar calculation for the low-$\xi$ component
  does not provide a useful constraint with $r_{\rm{low-}\xi} < 10^{23}
  ~\rm{cm} \sim 3 ~\rm{kpc}$, and, therefore, the location of the low-$\xi$
  component is not well-determined compared to that of the high-$\xi$
  component.  If the low-$\xi$ component lies beyond the high-$\xi$ component
  relative to the central continuum source, the difference in the measured
  bulk velocity shifts of the two components might indicate that the
  high-$\xi$ gas is the base of an accelerating outflow.  If, on the other
  hand, the low-$\xi$ absorber is indeed spatially coincident with the torus,
  in which case $r_{\rm{low-}\xi}$ is likely to be approximately the location
  of the broad-line region ($r_{\rm{low-}\xi} \sim r_{\rm{BLR}} \sim 1
  ~\rm{pc}$), the medium is decelerating as a function of radius.  Such a
  behavior has been observed in the UV spectrum of the Seyfert~1 galaxy
  NGC~4151 (Crenshaw et al.\ \cite{crenshaw00}).

  As briefly mentioned earlier, the source during the present {\it XMM-Newton}
  observation was in an unusually low state.  However, since the estimated
  location and density of the absorbing medium are such that the gas does not
  respond immediately to the observed continuum radiation (i.e., low density
  gas at large distances), the effect on the physical state of the absorbing
  medium by a variable illuminating source is not clear.  It will be useful to
  re-observe \object{IRAS~13349+2438} during a substantially brighter state to
  see whether the spectrum exhibits any dramatic changes in the absorption
  structures, and specifically if the oxygen absorption edges detected in the
  {\it ROSAT} and {\it ASCA} data (Brandt, Fabian, \& Pounds \cite{brandt96};
  Brandt et al.\ \cite{brandt97}) are really present during a different state.
  A detailed comparison of the UV and X-ray absorption spectra will also be
  interesting, particularly for identifying discrete kinematic components, as
  well as for constraining the global characteristics and dynamics of the
  absorbing medium.

\section{Comparisons with Other AGNs}

  The absorption spectrum of \object{IRAS~13349+2438} is qualitatively similar
  to those obtained with the {\it Chandra} transmission grating observations
  of the Seyfert 1 galaxies NGC 5548 and NGC 3783, which show narrow
  absorption lines blue-shifted by several hundred km~s$^{-1}$. (Kaastra et
  al.\ \cite{kaastra00}; Kaspi et al.\ \cite{kaspi00}).  The derived ion
  column densities in these sources, as well as in \object{IRAS~13349+2438},
  are in the range $N_i \sim 10^{16} - 10^{17} ~\rm{cm}^{-2}$, and are not
  high enough to produce observable absorption edges.

  Conceptually, the low-ionization component observed in
  \object{IRAS~13349+2438} is similar to the ``lukewarm absorber'' that
  explains both the observed optical and X-ray attenuation in NGC 3227
  (Kraemer et al.\ \cite{kraemer00}).  The spectroscopic signatures, however,
  are very different.  In particular, for the column densities derived from
  the \object{IRAS~13349+2438} data, the absorption features are dominated by
  discrete resonance line transitions, mainly in He-like ions and M-shell
  iron, and not by photoelectric edges as in the model of Kraemer et al.\
  (\cite{kraemer00}).

  As demonstrated in our detailed spectral analysis of
  \object{IRAS~13349+2438}, the UTA feature is potentially a powerful
  diagnostic tool for probing cool absorbing material using high-resolution
  X-ray observations.  If the low-ionization component is indeed associated
  with the dusty torus as the derived column density suggests, this feature
  should be detectable in other AGNs where the line-of-sight is partially
  obscured by the torus.

\begin{acknowledgements}
  We thank the referees Niel Brandt and Sarah Gallagher for constructive
  comments that helped improve the quality of the manuscript.  The Columbia
  University team is supported by NASA.  The Laboratory for Space Research
  Utrecht is supported financially by the Netherlands Organization for
  Scientific Research (NWO).  Work at LLNL was performed under the auspices of
  the U.S.\ Department of Energy by the University of California Lawrence
  Livermore National Laboratory under contract No.\ W-7405-Eng-48.
\end{acknowledgements}

\end{document}